\documentclass[floatfix,twocolumn]{aastex63}

\newcommand{\be}{\begin{eqnarray} }
\newcommand{\ee}{ \end{eqnarray} }

\shorttitle{A single pulse study of PSR~J1022+1001}
\shortauthors{Feng et al.}
\graphicspath{{./}{figures/}}
\begin{document}

\title{A single pulse study of PSR~J1022+1001 using the FAST radio telescope}

\correspondingauthor{Yi Feng}
\email{yifeng@nao.cas.cn}
\author{Yi Feng}
\affil{CAS Key Laboratory of FAST, National Astronomical Observatories, Chinese Academy of  Sciences, Beijing 100101, People's Republic of China}
\affil{University of Chinese Academy of Sciences, Beijing 100049, People's Republic of China}
\affil{CSIRO Astronomy and Space Science, PO Box 76, Epping, NSW 1710, Australia}

\author{G. Hobbs}
\affil{CSIRO Astronomy and Space Science, PO Box 76, Epping, NSW 1710, Australia}

\author{D. Li}
\affil{CAS Key Laboratory of FAST, National Astronomical Observatories, Chinese Academy of  Sciences, Beijing 100101, People's Republic of China}
\affil{University of Chinese Academy of Sciences, Beijing 100049, People's Republic of China}
\affil{NAOC-UKZN Computational Astrophysics Centre, University of KwaZulu-Natal, Durban 4000, South Africa}

\author{S. Dai}
\affil{CSIRO Astronomy and Space Science, PO Box 76, Epping, NSW 1710, Australia}

\author{W.~W.~Zhu}
\affil{CAS Key Laboratory of FAST, National Astronomical Observatories, Chinese Academy of  Sciences, Beijing 100101, People's Republic of China}

\author{Y.~L.~Yue}
\affil{CAS Key Laboratory of FAST, National Astronomical Observatories, Chinese Academy of  Sciences, Beijing 100101, People's Republic of China}

\author{P.~Wang}
\affil{CAS Key Laboratory of FAST, National Astronomical Observatories, Chinese Academy of  Sciences, Beijing 100101, People's Republic of China}

\author{S.-B. Zhang}
\affil{Purple Mountain Observatory, Chinese Academy of Sciences, Nanjing 210008, China}
\affil{University of Chinese Academy of Sciences, Beijing 100049, People's Republic of China}
\affil{CSIRO Astronomy and Space Science, PO Box 76, Epping, NSW 1710, Australia}

\author{L.~Qian}
\affil{CAS Key Laboratory of FAST, National Astronomical Observatories, Chinese Academy of  Sciences, Beijing 100101, People's Republic of China}

\author{L.~Zhang}
\affil{CAS Key Laboratory of FAST, National Astronomical Observatories, Chinese Academy of  Sciences, Beijing 100101, People's Republic of China}
\affil{University of Chinese Academy of Sciences, Beijing 100049, People's Republic of China}
\affil{CSIRO Astronomy and Space Science, PO Box 76, Epping, NSW 1710, Australia}

\author{S.~Q.~Wang}
\affil{Xinjiang Astronomical Observatory, 150, Science-1 Street, Urumqi, 830011 Xinjiang, China}

\author{C.~C.~Miao}
\affil{CAS Key Laboratory of FAST, National Astronomical Observatories, Chinese Academy of  Sciences, Beijing 100101, People's Republic of China}
\affil{University of Chinese Academy of Sciences, Beijing 100049, People's Republic of China}

\author{M.~Yuan}
\affil{CAS Key Laboratory of FAST, National Astronomical Observatories, Chinese Academy of  Sciences, Beijing 100101, People's Republic of China}
\affil{University of Chinese Academy of Sciences, Beijing 100049, People's Republic of China}

\author{Y.-K. Zhang}
\affil{CAS Key Laboratory of FAST, National Astronomical Observatories, Chinese Academy of  Sciences, Beijing 100101, People's Republic of China}
\affil{University of Chinese Academy of Sciences, Beijing 100049, People's Republic of China}



\begin{abstract}

Using the Five-hundred-meter Aperture Spherical radio Telescope (FAST), we have recorded $\sim 10^5$ single pulses from PSR~J1022+1001.  We studied the polarization properties, their energy distribution and their times of arrival.
This is only possible with the high sensitivity available using FAST. 
There is no indication that PSR~J1022+1001 exhibits giant pulse, nulling  or traditional mode changing phenomena.   
%
%The energy in the leading and trailing components of
%the integrated profile are correlated.
%
%
%The degrees of both linear and circular polarization increase with the %pulse peak flux density.
%
The energy in the leading and trailing components of the integrated profile is shown to be correlated. The degree of both linear and circular polarization increases with the pulse flux density for individual pulses. Our data indicates that pulse jitter leads to an excess noise in the timing residuals of 67\,ns when scaled to one hour, which is consistent with Liu et al. (2015).  We have unsuccessfully trialled various methods to improve timing precision through the selection of specific single pulses. Our work demonstrates that FAST can detect individual pulses from pulsars that are observed in order to detect and study  gravitational waves. This capability enables detailed studies, and parameterisation, of the noise processes that affect the sensitivity of a pulsar timing array.

%Although none of the methods work for this pulsar, FAST will provide the opportunity to improve the sensitivity of pulsar timing array observations and for other key pulsar-related research.

%
%Finally, selecting single pulses in different signal to noise ratio ranges or according to component energy cannot improve timing precision for PSR~J1022$+$1001.    
%The discovery of swift mode changes is relevant to pulsar timing array activities on current and future telescopes. We show that the timing using only one emission state does not improve the timing precision for this pulsar and discuss possible explanations for this. 
%We propose a criterion of timing precision improvement with a certain emission state of swift mode changes.   

\end{abstract}



\section{Introduction} \label{sec:intro}
Radio pulsars are known to exhibit a profusion of emission phenomena. For instance, mode changing (e.g. \citealt{1982ApJ...258..776B}, \citealt{2007MNRAS.377.1383W}), nulling (e.g.  \citealt{1970Natur.228...42B}, \citealt{2007MNRAS.377.1383W}), sub-pulse drifting (e.g. \citealt{1968Natur.220..231D}, \citealt{2006A&A...445..243W}) and giant pulse (e.g. \citealt{1968Sci...162.1481S}, \citealt{2003Natur.422..141H}) phenomena have been observed in  normal pulsars. 
The profile shapes of individual pulses vary significantly (e.g. \citealt{2001ApJ...546..394J}, \citealt{2012ApJ...761...64S}), but the summation of a large number of pulses usually leads to a stable profile. Mode changing is a discontinuous change where
the mean pulse profile abruptly changes between two (or sometimes
more) quasi-stable states.  Nulling is where individual pulses in the pulse train are undetectable.  
Individual pulses can be tens or hundreds of times brighter than the average, which is known as giant pulse phenomenon. Giant pulses are much narrower than the average profile and have durations that range from nanoseconds to microseconds and follow a power-law energy distribution. With this definition, giant pulses have been detected in seven pulsars, including two young pulsars and five millisecond pulsars (e.g. \citealt{1968Sci...162.1481S}, \citealt{1996ApJ...457L..81C}, \citealt{2003ApJ...590L..95J}). 
Mode changing and nulling are thought to be linked and are caused by changes in the pulsar magnetosphere \citep{2010Sci...329..408L,2019MNRAS.485.3230S}.

Mode changing, nulling and sub-pulse drifting phenomena are relatively rare in millisecond pulsars (MSPs). Mode changing has been observed in only one MSP \citep{2018ApJ...867L...2M}, nulling has not yet been reported and sub-pulse drifting has been observed in only three MSPs (\citealt{2003A&A...407..273E}, \citealt{2015MNRAS.449.1158L}).  This cannot be fully explained by selection biases caused by a lack of well studied single pulses from MSPs (e.g. \citealt{2001ApJ...546..394J}, \citealt{2012ApJ...761...64S}, \citealt{2014MNRAS.441.3148O}, \citealt{2016MNRAS.463.3239L}). 

\begin{figure}
%\centering
%\includegraphics[width=80mm]{stokes.png}
\includegraphics[width=8.5cm]{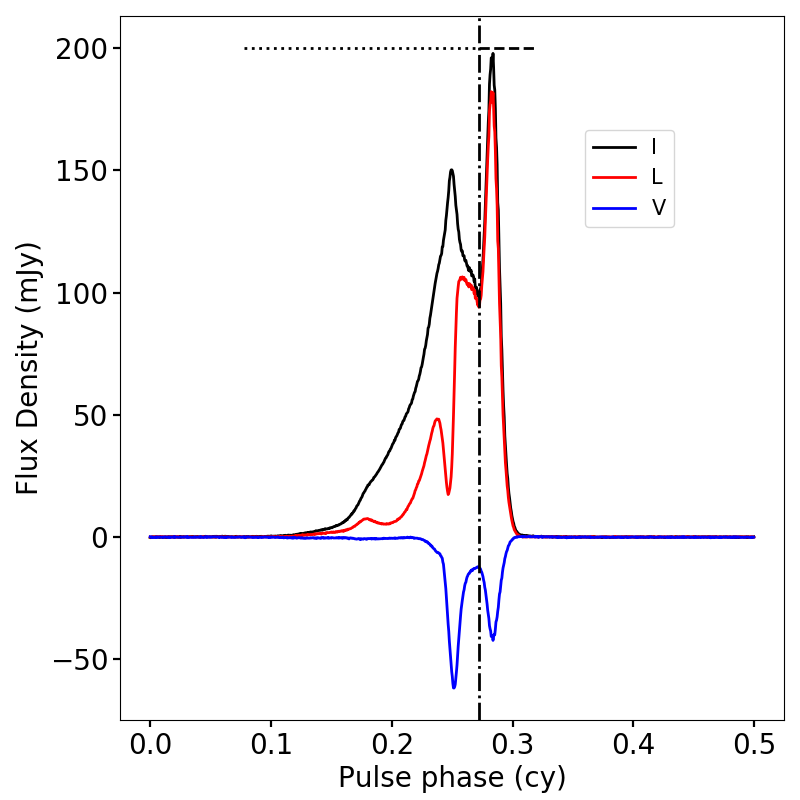}
\caption{Averaged polarization profile of PSR~J1022$+$1001. The black, red and blue lines represent Stokes I,
the linear polarization and Stokes V, respectively. We define the phase ranges identifying the leading and trailing components using the horizontal and vertical lines (chosen to agree with \citealt{2015MNRAS.449.1158L} and to be the minimum between the two peaks in the profile). The on-pulse region is taken to be the range delimited by the leading and trailing components.}
\label{profile}
\end{figure}

In this paper, we carried out a single pulse study of PSR~J1022+1001. We have recorded over $\sim 10^5$ single pulses from  PSR~J1022+1001 with the Five-hundred-meter Aperture Spherical radio Telescope \citep[FAST, ][]{2011IJMPD..20..989N, li16}.  PSR~J1022$+$1001 has a pulse period of 16.45\,ms and a dispersion measure (DM) of 10.2\,$\rm {cm^{-3}}\ pc$. The scintillation bandwidth and time-scale for this pulsar are 65\,MHz and 2334\,s respectively at 1400\,MHz \citep{2013MNRAS.429.2161K}. The pulsar is in a binary system with a 7.8-d orbital period with a white dwarf companion. The average pulse profile obtained from our FAST observation is shown in Figure~\ref{profile}. The integrated pulse profile of PSR~J1022+1001 consists of two components in the 20-cm band with a highly linearly polarized trailing component.

The stability of the integrated profile for PSR~J1022+1001 has been studied in detail.  \cite{1999ApJ...520..324K} showed that the amplitude ratio of the two components evolves on short time-scales (tens of minutes).  \cite{2004MNRAS.355..941H} argued that profile instability could also arise from improper polarization calibration because of an imperfect receiver model. \cite{2015MNRAS.449.1158L} showed that improper polarization calibration and diffractive scintillation cannot be the sole reason for the observed profile instability. \cite{2017ChA&A..41..495S} argued that profile instability is mainly caused by the evolution of the pulse profile with frequency combined with the interstellar scintillation. Recently with long term observations at Effelsberg Radio Telescope, \cite{2020MNRAS.tmp.2964P} found that the pulse shape variations cannot be fully accounted for by
instrumental and propagation effects and suggested additional intrinsic effects as the origin for the variation. 

%Three distinct classes of PSR~J1022+1001 single pulses have been detected using our observations, which we attribute to three  emission states. The averaged pulse profiles of the three emission states have different shapes and remain stable with time, which is similar to the mode changing phenomenon. However, PSR~J1022+1001 transits from a certain emission state to another in just seven spin periods, which is much shorter than those of normal mode-changing pulsars. Therefore the ``new'' emission phenomenon is tentatively called swift mode change.  

Our observations and data processing procedures are described in Section~\ref{sec:data}. Our results are presented in Section~\ref{sec:results}.  
%In Section~\ref{sec:results} we first measure jitter noise level and compare the results with published results. Then we show residual-time, residual-S/N relations and classify the single pulses according to the relations. Based on the classification of single pulses, we define and quantify two types of jitter noise. Timing using a certain class of single pulses is tested in Section~\ref{sec:results}. 
We conclude and discuss our results in Section~\ref{sec:conclusion}.

\section{Observation and Data processing} \label{sec:data}
%\subsection{observations } \label{sec:observations}
For our analysis, we selected one of the brightest observations of PSR~J1022$+$1001 from a FAST early science project (PID~3005).
We obtained $7.3\times10^{4}$ single pulses in a 20 minute observation on MJD~58749. 
The observation was conducted using the central beam of the 19-beam receiver \citep{2019SCPMA..6259502J}. The 19-beam receiver, with the frequency range between 1050 and 1450\,MHz, provides two data streams (one for each hand of linear polarization). The data streams are processed with the Reconfigurable Open Architecture Computing Hardware–version 2 (ROACH2) signal processor \citep{2019SCPMA..6259502J}. The output data files are recorded as 8 bit-sampled search mode PSRFITS \citep{2004PASA...21..302H} files with 4096 frequency channels and with 49.152\,$\mu$s time resolution. 

%We recorded $7.3\times10^4$ pulses for PSR~J1022$+$1001 over a 20 minute integration on MJD 58749. 
We used the DSPSR software package \citep{2011PASA...28....1V} to extract individual pulses. The ``-K'' option of the DSPSR software package was used to remove inter-channel dispersion delays. Polarization calibration was achieved by correcting for the differential gain and phase between the receptors through separate measurements of a noise diode signal injected at an angle of $45^{\circ}$ from the linear receptors. We determined a rotation measure (RM) of 2.9$\pm$0.2\,rad\,m$^{-2}$ from our data and the profiles are RM-corrected. The observations were then flux density
calibrated using observations of 3C 286 \citep{1977A&A....61...99B}. 
To excise radio frequency interference (RFI), we used the PSRCHIVE software package \citep{2004PASA...21..302H} to median filter each single pulse in the frequency domain. Band-averaged analytical templates were made using the profile obtained when an entire observation was folded. We used the PSRCHIVE  and TEMPO2 \citep{hobbs2006} software packages to determine pulse times of arrival (ToAs) and timing residuals for each single pulse using the timing model from \cite{2020PASA...37...20K}. 

Our calibration procedure produces an integrated pulse profile, shown in Figure~\ref{profile}, that agrees with previously published profiles (for instance, \citealt{2015MNRAS.449.3223D}).

\section{Results} \label{sec:results}

We first present our results from analysing the single pulse energy distributions and our search for correlations between the leading and trailing profile components.   We then search for the presence of giant pulse, nulling or traditional mode changing phenomena.  We study the polarization properties of the single pulses. Finally, we measure the jitter noise and trial various methods to improve the timing precision by only selecting a sub-set of the single pulses.

\subsection{Single pulse statistics}

\begin{figure}
%\centering
%\includegraphics[width=80mm]{stokes.png}
\includegraphics[width=8.5cm]{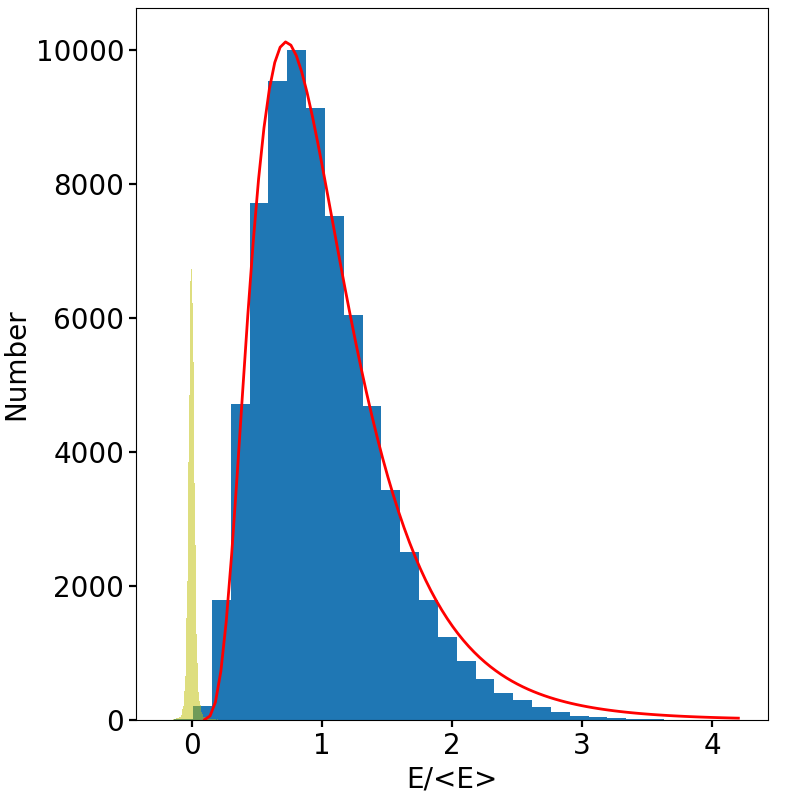}
\caption{Pulse energy histograms for the off-pulse region (yellow) and on-pulse region (blue). The pulse energies are normalised by the mean on-pulse energy. The red line is a log-normal distribution that has been fitted to the on-pulse energy histogram.}
\label{energy}
\end{figure}

We define the energy of each single pulse (in non-physical units) to be its integrated
flux density. For each pulse, this is determined by measuring the area across the on-pulse region (see ranges in Figure~\ref{profile}). We also defined an off-pulse window that was used as a control sample to assess the statistics of the noise in our measurements. The off–pulse window was chosen to have the same width as the on-pulse region.  We normalised energy of each pulse by the mean value in the on-pulse region $\left\langle E \right\rangle$.  The results are shown in Figure~\ref{energy}. The yellow histogram shows the off-pulse energy distribution and the noise in the off-pulse region follows the expected normal distribution. The blue histogram shows the on-pulse energy distribution, which is well fit (see red line in Figure~\ref{energy}) by a log-normal distribution of $N(E)  = \frac{A}{E} \exp(-\frac{(\ln E-\mu)^2}{2\sigma^2})$ where $\mu = -0.06\pm0.01$ and $\sigma = 0.51\pm0.01$ and A is normalization factor.

\begin{figure}
%\centering
%\includegraphics[width=80mm]{stokes.png}
\includegraphics[width=8.5cm]{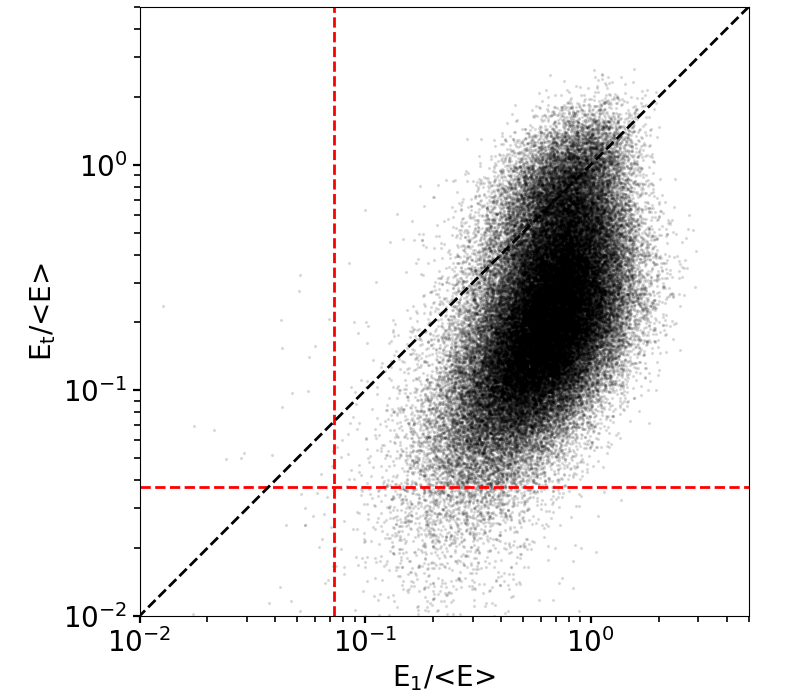}
\caption{Correlation of the leading component pulse energy and trailing component pulse energy. The pulse energies are normalised by the mean on-pulse energy. The vertical and horizontal red dashed lines represent the $3\,\sigma$ noise levels of the leading and trailing component respectively. The black dashed line represents $E_t = E_l$.}
\label{energy_cor}
\end{figure}

For each single pulse, we also measured the leading component pulse energy, $E_{l}$, using the area across the leading component region, as defined using the dotted horizontal line in Figure~\ref{profile}. Similarly, we measured trailing component pulse energy (defined by the dashed line in the Figure). Figure~\ref{energy_cor} shows the correlation between the leading component pulse energy, $E_{l}$, and the trailing component pulse energy, $E_{t}$. The Pearson product-moment correlation coefficient of $E_{l}$ and $E_{t}$ is 0.38. The points do not fall on the $E_t = E_l$ line. The leading component is much wider in general, and, as a result, the leading component has more energy than that of the trailing component in most cases. 

\cite{2015MNRAS.449.1158L} showed that the occurrence of subpulses (individual components of the single pulses) in the leading and trailing components of the PSR~J1022+1001 profile is correlated.  Their work was based on 14000 subpulses obtained over 35-h, which is only 0.2\% of the total number of pulses expected during that time period.   They have therefore only selected the brightest pulses for their analysis. We note from our correlation that if the leading component energy is large, then the trailing component energy also tends to be large.  This agrees with the results in \cite{2015MNRAS.449.1158L}, but we note from our data that the components are only partially correlated.

\subsection{Nulling and giant pulses}

As shown in Figure~\ref{energy} the energy distribution can be modelled used a single log-normal distribution. We would expect the energy distribution of nulling pulses is similar to that of the off-pulse region (the yellow histogram).  However, we see no evidence of this with our data set. If we define nulling pulses to have on-pulse energy smaller than three times the standard deviation of the off-pulse energy, then this gives an upper limit of nulling fraction of 0.06\%. 

 The pulse energy for normal pulses typically follows a log-normal or normal distribution. In comparison, the pulse energy for giant pulses generally follows a power-law distribution, e.g. the Crab pulsar~\citep{1968Sci...162.1481S}, PSR B1937+21~\citep{1996ApJ...457L..81C} and PSR B0540$-$69~\citep{2003ApJ...590L..95J}. In our data, the brightest pulse is only approximately four times the average. We therefore have no evidence that PSR~J1022+1001 exhibits the giant pulse phenomenon. 

\subsection{Moding-changing}

Mode-changing behaviour has previously only been seen in one other millisecond pulsar (PSR~B1957+20). The switching cadence between the modes for that pulsar is $\sim$1000 pulses. This is similar to the mode switching time-scales found in normal, non-millisecond pulsars (but note these time scales are much shorter than the pulse variations identified for intermittent pulsars; e.g. \citealt{2020ApJ...897....8W}). 

\begin{figure}
%\epsscale{1.2} %1.2
\begin{center}
%\plotone{1022_stack.png}
\includegraphics[width=0.5\textwidth]{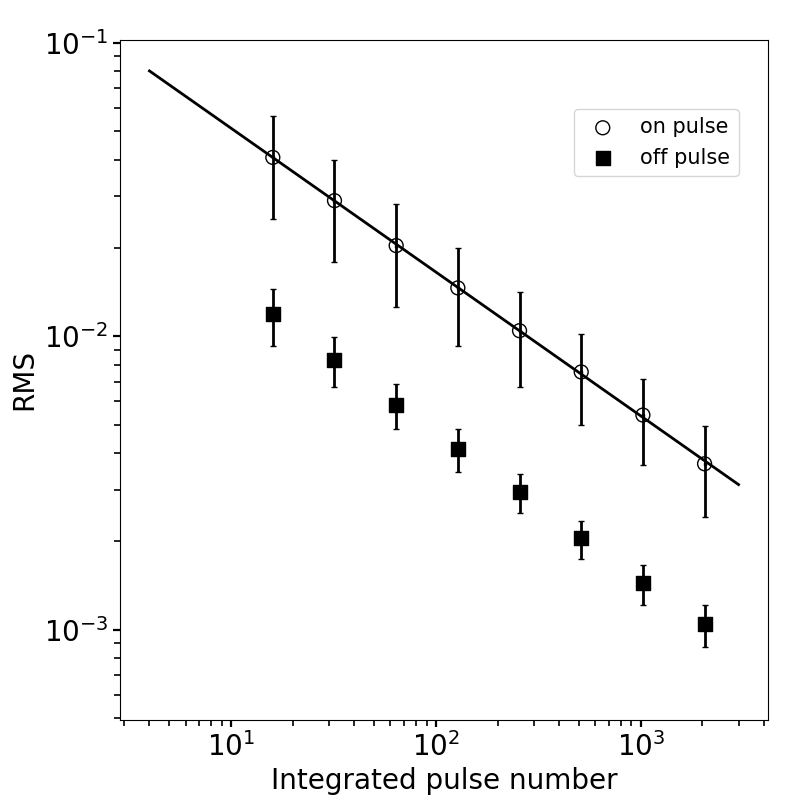}
\caption{Root mean square (RMS) values obtained from the difference between the profile formed for a given number of pulses and a template obtained using all available pulses.  The open circles represent the RMS values obtained in the on-pulse region. The solid black squares are obtained from the off-pulse region. The solid black line is the best-fitting model for the on-pulse data.}
\label{testmode}
%D:\projects\project_jitter\toPub\ana\psrfits\1022\stack_multi.py
\end{center}
\end{figure}

To determine whether PSR~J1022+1001 shows mode-changing, we averaged different numbers of individual pulses together to form pulse profiles. We then used the PSRCHIVE software package to compare the averaged pulse profiles with the template obtained by summing all the available data.   We subtracted each profile from the phase aligned template and determined the root mean square (RMS) values of the difference in the on-pulse region and in the off-pulse region.  These RMS values are plotted in Figure~\ref{testmode}.   We fitted a power law, $\mathrm{RMS}(N) = A N^{\beta}$ (where $N$ is the number of pulses that have been averaged), to the on-pulse RMS values and obtained that $\beta = -0.49 \pm 0.01$. There is no indication from this Figure that PSR~J1022+1001 undergoes profile changes on a time scale between $\sim 10$ and $\sim 2000$ pulses. 

\begin{figure}
%\centering
%\includegraphics[width=80mm]{stokes.png}
\includegraphics[width=8.5cm]{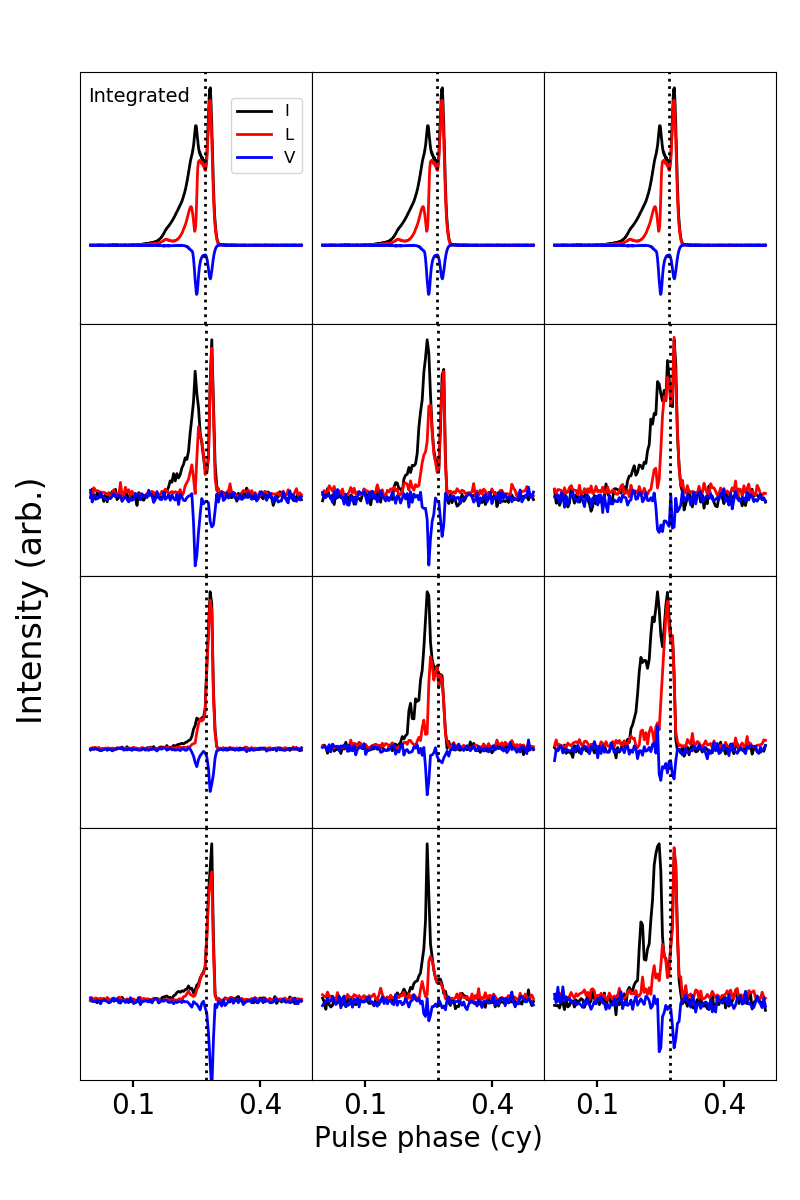}
\caption{Single pulses from PSR~J1022$+$1001. The first row shows the integrated pulse profile. The black, red and blue lines represent Stokes I, the linear polarization and Stokes V, respectively. The vertical dashed line separates the leading component and the trailing component. The left column shows three single pulses which have a brighter trailing component than the leading component. The middle column shows three single pulses which have a brighter leading component. The right column shows three single pulses which have multiple peaks.}
\label{shape}
\end{figure}

\subsection{Single pulse polarization}

The single pulses have a large variety of profile shapes. In Figure~\ref{shape} we show polarization profiles for a few, representative, single pulses.  The top row shows three copies of the integrated pulse profile.  For the remaining nine panels, the left column contains three single pulses, which have a brighter trailing component than the leading component. The middle column shows three single pulses which have a brighter leading component. The right column shows three single pulses which have multiple peaks. When the leading component dominates, the trailing component varies from being almost as bright as the leading component, to being undetectable.  The same is true when the trailing component dominates. We note that the trailing component is always highly linearly polarized, however, the leading component has a varying degree of linear polarization.  

\begin{figure}
%\centering
%\includegraphics[width=80mm]{stokes.png}
\includegraphics[width=8.5cm]{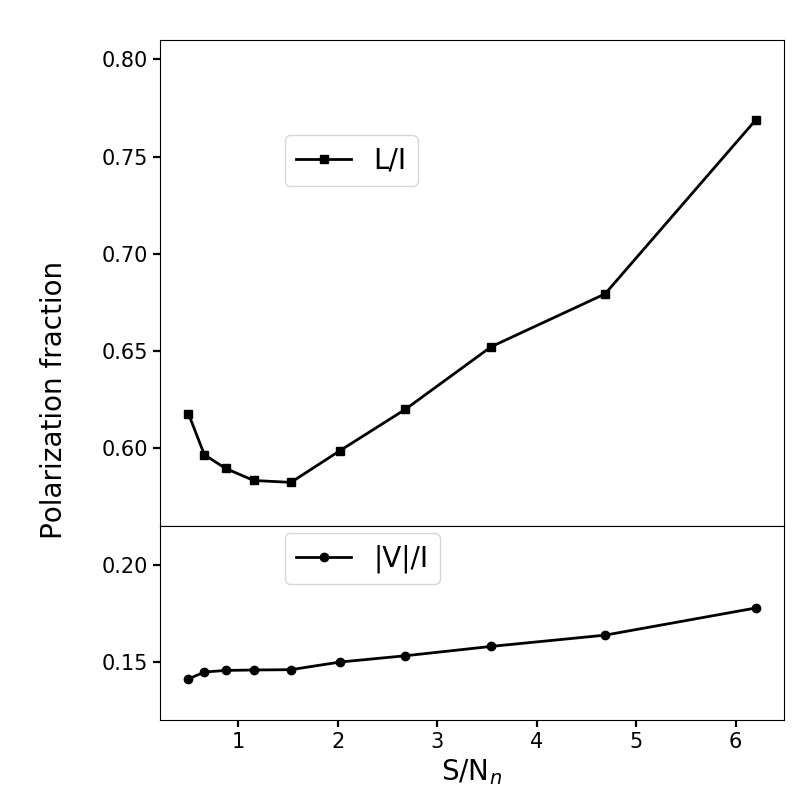}
\caption{Polarization fractions of linear (top) and circular (bottom) component as a function of S/N$_{n}$.}
\label{pol}
\end{figure}

In order to determine the degree of correlation between the signal-to-noise (S/N) ratio for each pulse and its polarization fraction, we defined the degree of linear polarization 
as ($\Sigma_{i}\sqrt{Q_i^{2}+U_i^{2}}$)/($\Sigma_{i}I_i$)
and that of circular polarization as ($\Sigma_{i}|V_i|$)/($\Sigma_{i}I_i$), where $Q,U,V$ are the Stokes parameters and the summation is over the phase bins in the on-pulse region. We determined the signal to noise ratio (S/N) as the integrated power in the pulse profile divided by the noise in the baseline. We then normalised these S/N values by dividing by the mean S/N value (giving $\mathrm{S/N}_n$).    Finally we divided the single pulses in $\mathrm{S/N}_{n}$ ranges in Table~\ref{tab:snr}.
%: $[0.0,0.50)$; $[0.50,0.66)$; $[0.66,0.87)$; $[0.87,1.16)$;
%$[1.16,1.53)$; $[1.53,2.03)$; $[2.03,2.68)$; $[2.68,3.54)$; $[3.54,4.69)$and $[4.69,6.2)$. 
Figure~\ref{pol} shows the relation between $\mathrm{S/N}_{n}$ and 
the degrees of linear and circular polarization.
The degrees of both linear and circular polarization increases with $\mathrm{S/N}_{n}$. This agrees with similar dependencies
observed in two other MSPs (\citealt{2014MNRAS.441.3148O}, \citealt{2016MNRAS.463.3239L}).   We note an apparent increase in the linear polarization fraction at low S/N.  This is also seen in the results presented by \cite{2014MNRAS.441.3148O} for PSR~J0437$-$4715. Using Equation (22) in \cite{askap20}, we calculated uncertainties on the linear polarization fraction to be smaller than 0.3\% for all points in the upper panel of Figure~\ref{pol}, thus this may represent some unexpected property of the low S/N pulses. 

%The uncertainties in the polarization fraction for an individual pulse with low S/N is clearly large, but we have summed thousands of pulses together when forming these results.  To date, we are unable to determine whether this is a bias in the polarization fraction determination or whether this represents some unexpected property of the low S/N pulses. 

\begin{table}
\caption{$\mathrm{S/N}_{n}$ ranges used for Figure~\ref{pol}.}
\begin{tabular}{ c  c  c }
\hline
Included range	& Number of pulses	 & Fraction of pulses \\
\hline
$0<\mathrm{S/N}_{n}<0.50$ & 18533	 & 0.2539 \\
$0.50<\mathrm{S/N}_{n}<0.66$ & 11428	 & 0.1565 \\
$0.66<\mathrm{S/N}_{n}<0.87$ & 10923	 & 0.1496 \\
$0.87<\mathrm{S/N}_{n}<1.16$ & 9346	 & 0.1280\\
$1.16<\mathrm{S/N}_{n}<1.53$ & 8705	 & 0.1192\\
$1.53<\mathrm{S/N}_{n}<2.03$ & 7359	 & 0.1008 \\
$2.03<\mathrm{S/N}_{n}<2.68$ & 4562	 & 0.0625 \\
$2.68<\mathrm{S/N}_{n}<3.54$ & 1759	 & 0.0241 \\
$3.54<\mathrm{S/N}_{n}<4.69$ & 352	 & 0.0048 \\
$4.69<\mathrm{S/N}_{n}<6.2$ & 33	 & 0.0005 \\
\hline
\end{tabular}
\label{tab:snr}
\end{table}

\subsection{Pulse jitter}

PSR~J1022$+$1001 is observed by pulsar timing array projects (e.g. \citealt{2013CQGra..30v4007H}, \citealt{2013PASA...30...17M}, \citealt{2013CQGra..30v4010M}, \citealt{2013CQGra..30v4009K}), which have the primary goal of detecting ultra-low-frequency gravitational waves.  The sensitivity of such an array depends upon the scatter in the timing residuals.  

On short time-scales the timing residuals are primarily dominated by radiometer noise (a reasonable representation is given by the ToA uncertainties) and pulse jitter (e.g. \citealt{2011MNRAS.418.1258O}, \citealt{2012MNRAS.420..361L}, \citealt{2012ApJ...761...64S}, \citealt{2015MNRAS.449.1158L}, \citealt{2014MNRAS.443.1463S}, \citealt{2016ApJ...819..155L}).  
Following \cite{2014MNRAS.443.1463S}, we define the jitter noise level obtained after $N$ pulses have been averaged together, $\sigma_{\rm J}(N)$,  to be the quadrature difference between the root mean square (RMS) of the timing residuals and the radiometer noise
\be
\label{eqn:jit_equation}
\sigma^2_{\rm J}(N)  = \sigma_{\rm obs}^2(N)  - \sigma_{\rm radiometer}^2(N).
\ee
This assumes that all of the excess error in the arrival time measurements can be attributed to jitter noise. We used simulated data sets to obtain the radiometer noise. In these simulated data sets, we formed pulse profiles from the template obtained by summing all the available pulses and white noise such that the S/N of the simulated profile matched the S/N of the observed profile. 

\begin{figure}
%\epsscale{1.2} %1.2
\begin{center}
%\plotone{1022_stack.png}
\includegraphics[width=0.5\textwidth]{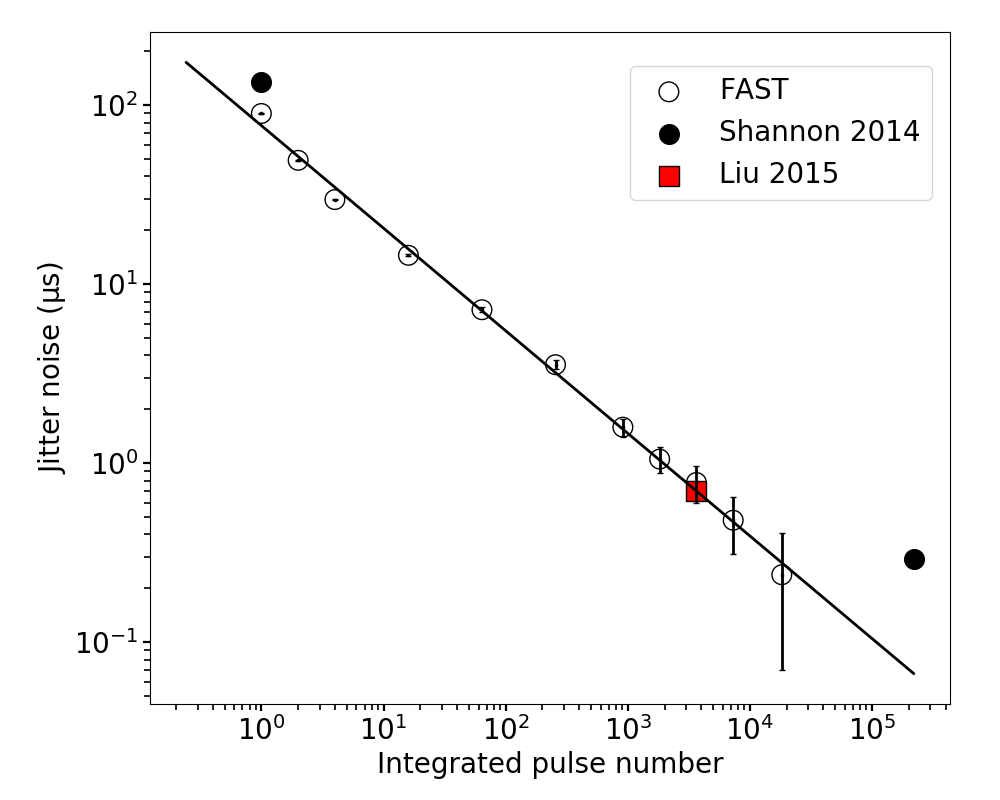}
\caption{Estimates of jitter noise in PSR~J1022$+$1001 from the FAST observation. Open circles are obtained with FAST using all single pulses. Solid black data points are the results in \cite{2014MNRAS.443.1463S}. The red square is the result from \cite{2015MNRAS.449.1158L}. 
The solid black line is the best-fitting model for the jitter noise measured with FAST.}
\label{1022_ryan}
%D:\projects\project_jitter\toPub\ana\psrfits\1022\stack_multi.py
\end{center}
\end{figure}

In Figure \ref{1022_ryan}, we compare estimates of the jitter noise in PSR~J1022$+$1001 from our FAST observation (open circles) with previous Parkes observations \citep{2014MNRAS.443.1463S}, and from Westerbork Synthesis Radio Telescope and Effelsberg 100-m Radio Telescope observations \citep{2015MNRAS.449.1158L}.  We fit a power law, $\sigma{_{\rm J}}(N) = A N^{\beta}$ to the jitter noise measured from our observations and obtain $\beta = -0.57 \pm 0.01$.  The jitter noise level when scaled to one hour is $67 \pm 9$\,ns. This value is significantly smaller than the value of $290 \pm 15$\,ns reported in \cite{2014MNRAS.443.1463S} (see black circles in Figure~\ref{1022_ryan}), but is consistent with \cite{2015MNRAS.449.1158L}, which reported that a jitter noise level of 700\,ns based on continuous 1-min integrations (red square in Figure~\ref{1022_ryan}).   The discrepancy between our results and \cite{2014MNRAS.443.1463S} may be caused by profile shape variations on hour-long time scales (which the \citealt{2014MNRAS.443.1463S} analysis probes). Such variations could be caused by the evolution of the pulse profile with frequency combined with the interstellar scintillation \citep{2017ChA&A..41..495S}. 

\subsection{Single pulse timing}
%\subsection{residual-time and residual-S/N relations}\label{sec:diagram}

\begin{figure*}
%\epsscale{1.2} %1.2
\begin{center}
%\plotone{1022_timing.png}
\includegraphics[width=16cm]{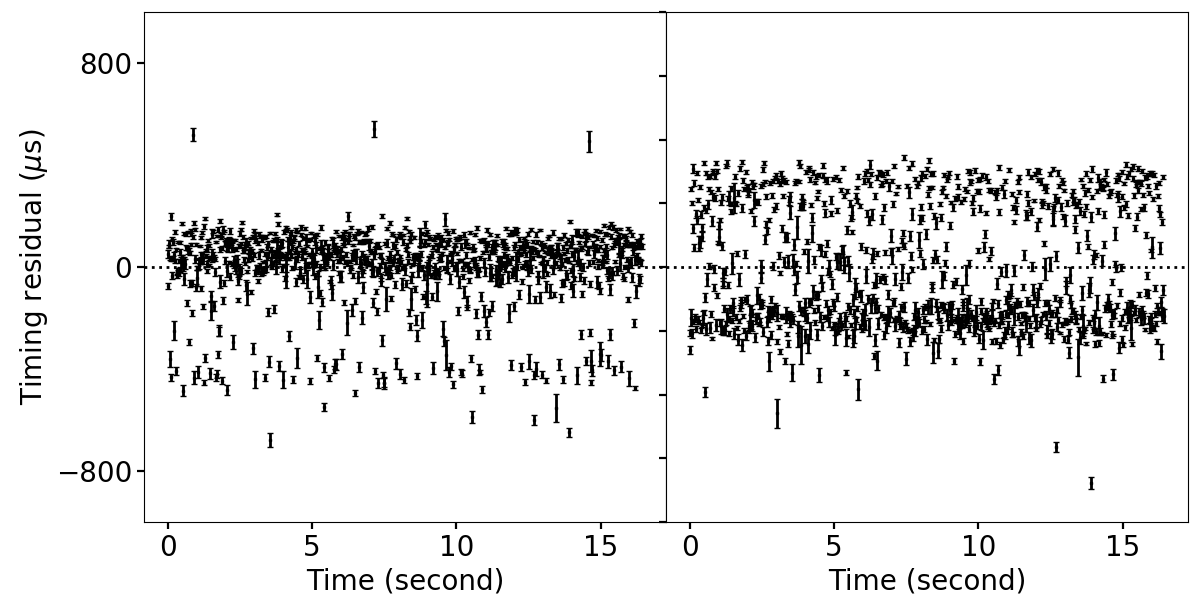}
\caption{Left and right panels show the first 1000 timing residuals using different templates. We used the standard template in the left-hand panel. In the right-hand panel we used a template obtained from the summation of the profiles that formed the most negative timing residuals in the left-hand panel.}
%For 1050 to 1450\,MHz frequency range, the points in green regions (148, 84 and 140 single pulses for class 1,2,3 respectively) are selected brightest single pulses of each class.}
\label{1022_timing}
%D:\projects\project_jitter\toPub\ana\psrfits\1022\timing.py
\end{center}
\end{figure*}

In the left-hand panel of Figure~\ref{1022_timing} we show the timing residuals obtained using our standard template for this pulsar. This provides the indication of three differing states. However, the timing residuals change with different templates.  In the right-hand panel we used a template obtained from the summation of the profiles that formed the most negative timing residuals in the left-hand panel.  We note that the number of residuals in each state (and the residual level of the states) varies between the two panels.

\begin{figure}
%\epsscale{1.2} %1.2
\begin{center}
%\plotone{1022_stack.png}
\includegraphics[width=0.5\textwidth]{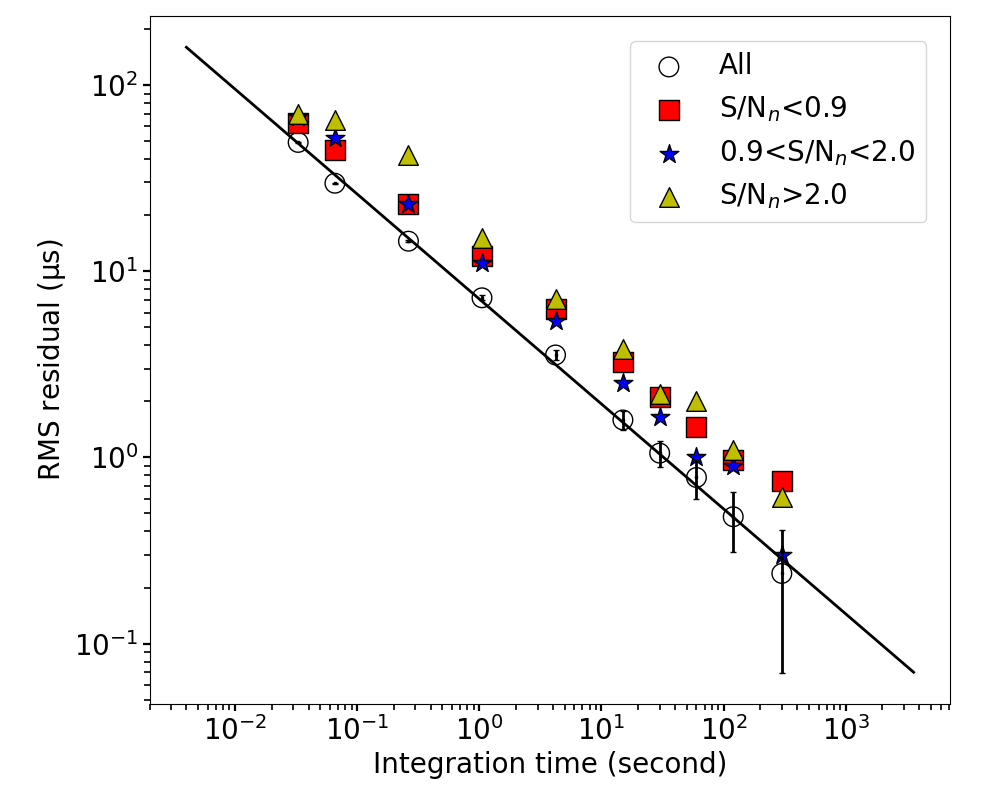}
\caption{RMS timing residuals as a function of integration time obtained using all available pulses (open circles) and sub-sets determined from the specific $\mathrm{S/N}_{n}$ ranges (red, blue and yellow symbols).   The solid black line is the best-fitting model for the RMS timing residual versus integration time determined using all the available pulses. }
\label{timing_bright}
%D:\projects\project_jitter\toPub\ana\psrfits\1022\stack_multi.py
\end{center}
\end{figure}

\cite{2014MNRAS.441.3148O} and \cite{2016MNRAS.463.3239L} trialled methods to improve the timing residuals by selecting pulses based on their S/N (for PSRs~J0437$-$4715 and J1713$+$0747 respectively).
Following these studies, we divided our single pulses into the following $\mathrm{S/N}_{n}$ ranges: $\mathrm{S/N}_{n}<0.9$, $0.9<\mathrm{S/N}_{n}<2.0$ and $\mathrm{S/N}_{n}>2.0$. We obtained a template by summing the single pulses from each $\mathrm{S/N}_{n}$ range, formed arrival times for the pulse and then timing residuals. The RMS timing residual is shown in Figure~\ref{timing_bright} as a function of integration time.  
The RMS residual values are higher for all of the $\mathrm{S/N}_{n}$ selections compared with simply using all of the available pulses. Although \cite{2014MNRAS.441.3148O} reported marginally reduction of RMS residual by rejecting $\sim 1\%$ of largest S/N single pulses, PSRs~J0437$-$4715 in \cite{2014MNRAS.441.3148O} and J1713$+$0747 in \cite{2016MNRAS.463.3239L} agree with our results.

\begin{figure}
%\epsscale{1.2} %1.2
\begin{center}
%\plotone{1022_stack.png}
\includegraphics[width=0.5\textwidth]{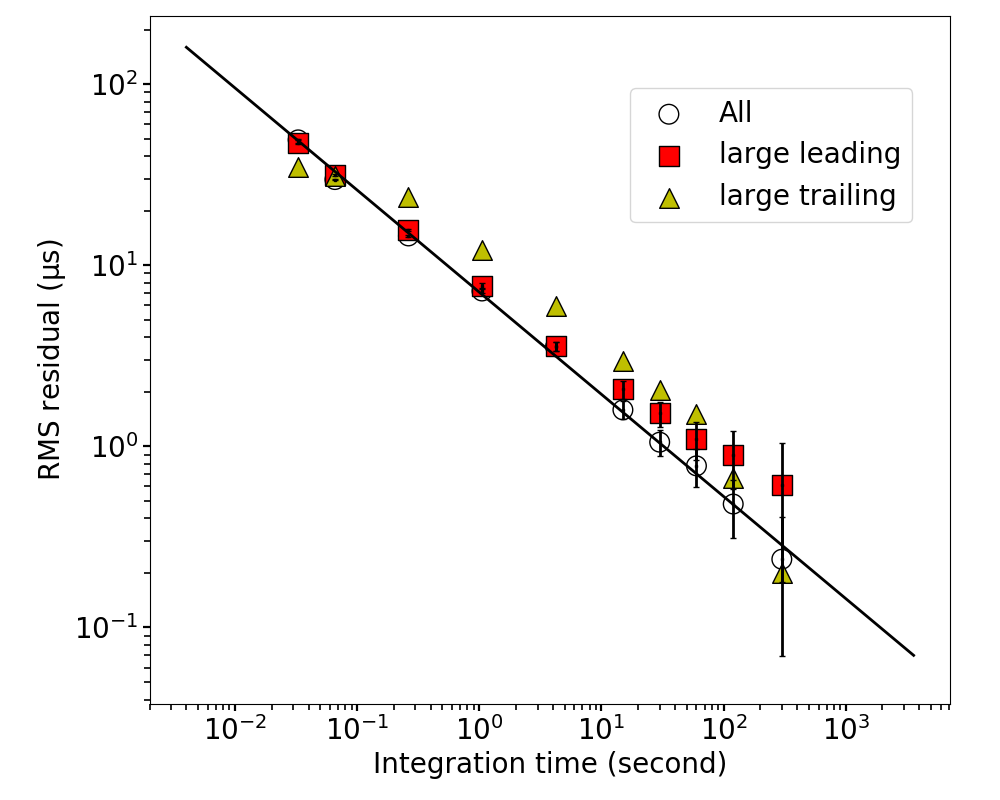}
\caption{RMS timing residuals as a function of integration time obtained using all pulses (open circles) and sub-sets obtained from the pulses in which the leading component is larger than the trailing component (red square) and vice versa (yellow triangle). 
The solid black line is the best-fitting model to the open circles.}
\label{timing_trailing}
%D:\projects\project_jitter\toPub\ana\psrfits\1022\stack_multi.py
\end{center}
\end{figure}

As PSR~J1022$+$1001 is a multi-component pulsar and the two components have, on average, comparable average energies, we also divided the single pulses according to the energy of the leading component and the trailing component: the single pulses which have larger leading component energy constitute the first class and the remainder constitute the second class. We followed the process above to form new templates from these samples and produced timing residuals.  The RMS timing residuals as a function of integration time are shown in Figure~\ref{timing_trailing}.  Again, we identify no improvement in the RMS residuals through this selection. The single pulses which have larger leading component energy consist of 91\% of all pulses, so the resulting RMS timing residuals are close to that of all pulses.

\section{Conclusions} \label{sec:conclusion}

We have  reported on the detection and analysis of $\sim10^5$ single pulses from PSR~J1022+1001.  This is only possible with the high sensitivity available using FAST. There is no indication that PSR~J1022+1001 exhibits giant pulse, nulling  or traditional mode changing phenomena.  The energy in the leading and trailing components of the integrated profile is shown to be correlated. The degree of both linear and circular polarization increases with the pulse flux density for individual pulses.

Jitter noise is a limiting source of noise at FAST and we have determined that the jitter noise for this pulsar, scaled to one hour, is 67\,ns. 
Selecting single pulses in different $\mathrm{S/N}_{n}$ ranges, or according to relative energy of the leading and trailing components, does not improve the timing precision achievable for PSR~J1022$+$1001.   

FAST provides us with the opportunity to study single pulses from millisecond pulsars in more detail than previously possible with the previous generation of radio telescopes.  In the near future, FAST will be able to provide a catalogue of single pulses from millisecond pulsars in the Northern Hemisphere and MeerKAT (and, in the near future, the Square Kilometre Array) will carry out similar work in the Southern Hemisphere. A detailed study of such single pulses may provide the opportunity to improve the sensitivity of pulsar timing array observations and for other key pulsar-related research, such as pulsar emission physics and testing theories of gravity.  

%Although we have only reported multiple emission states in one MSP (PSR~J1022+1001), this phenomenon may be common in MSPs with multiple components. For example, figures 11 and 17 in \cite{2014MNRAS.443.1463S} indicates signs of multiple emission states for PSRs~J1603$-$7202 and J2145$-$0750: the average profiles of the 100 brightest pulses have very different shapes from the corresponding integrated pulse profiles. The greater sensitivity of FAST, MeerKAT and the Square Kilometre Array will provide opportunities to find more MSPs with this phenomenon.

\acknowledgments
We would like to thank R. M. Shannon, R. N. Manchester, M. Bailes, W. van Straten and W. A. Coles for valuable discussions. We also thank the anonymous referee for providing constructive suggestions to improve the article.
This work is supported by National Key R$\&$D Program of China No. 2017YFA0402600, by the CAS International Partnership Program No.114-
A11KYSB20160008, by the CAS Strategic Priority Research Program No. XDB23000000, and by NSFC grant No. 11725313 and 11690024. YF is supported by China Scholarship Council. This work made use of the data from FAST (Five-hundred-meter Aperture Spherical radio Telescope).  FAST is a Chinese national mega-science facility, operated by National Astronomical Observatories, Chinese Academy of Sciences. 
\software{DSPSR (\citealt{2011PASA...28....1V}), PSRCHIVE (\citealt{2004PASA...21..302H}), TEMPO2 (\citealt{hobbs2006})}




\bibliography{1022}{}

\begin{thebibliography}{}
\expandafter\ifx\csname natexlab\endcsname\relax\def\natexlab#1{#1}\fi
\providecommand{\url}[1]{\href{#1}{#1}}
\providecommand{\dodoi}[1]{doi:~\href{http://doi.org/#1}{\nolinkurl{#1}}}
\providecommand{\doeprint}[1]{\href{http://ascl.net/#1}{\nolinkurl{http://ascl.net/#1}}}
\providecommand{\doarXiv}[1]{\href{https://arxiv.org/abs/#1}{\nolinkurl{https://arxiv.org/abs/#1}}}

\bibitem[{{Baars} {et~al.}(1977){Baars}, {Genzel}, {Pauliny-Toth}, \&
  {Witzel}}]{1977A&A....61...99B}
{Baars}, J.~W.~M., {Genzel}, R., {Pauliny-Toth}, I.~I.~K., \& {Witzel}, A.
  1977, \aap, 500, 135

\bibitem[{{Backer}(1970)}]{1970Natur.228...42B}
{Backer}, D.~C. 1970, \nat, 228, 42, \dodoi{10.1038/228042a0}

\bibitem[{{Bartel} {et~al.}(1982){Bartel}, {Morris}, {Sieber}, \&
  {Hankins}}]{1982ApJ...258..776B}
{Bartel}, N., {Morris}, D., {Sieber}, W., \& {Hankins}, T.~H. 1982, \apj, 258,
  776, \dodoi{10.1086/160125}

\bibitem[{{Cognard} {et~al.}(1996){Cognard}, {Shrauner}, {Taylor}, \&
  {Thorsett}}]{1996ApJ...457L..81C}
{Cognard}, I., {Shrauner}, J.~A., {Taylor}, J.~H., \& {Thorsett}, S.~E. 1996,
  \apjl, 457, L81, \dodoi{10.1086/309894}

\bibitem[{{Dai} {et~al.}(2015){Dai}, {Hobbs}, {Manchester}, {Kerr}, {Shannon},
  {van Straten}, {Mata}, {Bailes}, {Bhat}, {Burke-Spolaor}, {Coles},
  {Johnston}, {Keith}, {Levin}, {Os{\l}owski}, {Reardon}, {Ravi}, {Sarkissian},
  {Tiburzi}, {Toomey}, {Wang}, {Wang}, {Wen}, {Xu}, {Yan}, \&
  {Zhu}}]{2015MNRAS.449.3223D}
{Dai}, S., {Hobbs}, G., {Manchester}, R.~N., {et~al.} 2015, \mnras, 449, 3223,
  \dodoi{10.1093/mnras/stv508}

\bibitem[{{Day} {et~al.}(2020){Day}, {Deller}, {Shannon}, {Qiu}, {Bannister},
  {Bhandari}, {Ekers}, {Flynn}, {James}, {Macquart}, {Mahony}, {Phillips}, \&
  {Xavier Prochaska}}]{askap20}
{Day}, C.~K., {Deller}, A.~T., {Shannon}, R.~M., {et~al.} 2020, \mnras, 497,
  3335, \dodoi{10.1093/mnras/staa2138}

\bibitem[{{Drake} \& {Craft}(1968)}]{1968Natur.220..231D}
{Drake}, F.~D., \& {Craft}, H.~D. 1968, \nat, 220, 231,
  \dodoi{10.1038/220231a0}

\bibitem[{{Edwards} \& {Stappers}(2003)}]{2003A&A...407..273E}
{Edwards}, R.~T., \& {Stappers}, B.~W. 2003, \aap, 407, 273,
  \dodoi{10.1051/0004-6361:20030716}

\bibitem[{{Hankins} {et~al.}(2003){Hankins}, {Kern}, {Weatherall}, \&
  {Eilek}}]{2003Natur.422..141H}
{Hankins}, T.~H., {Kern}, J.~S., {Weatherall}, J.~C., \& {Eilek}, J.~A. 2003,
  \nat, 422, 141, \dodoi{10.1038/nature01477}

\bibitem[{{Hobbs}(2013)}]{2013CQGra..30v4007H}
{Hobbs}, G. 2013, Classical and Quantum Gravity, 30, 224007,
  \dodoi{10.1088/0264-9381/30/22/224007}

\bibitem[{{Hobbs} {et~al.}(2006){Hobbs}, {Edwards}, \&
  {Manchester}}]{hobbs2006}
{Hobbs}, G.~B., {Edwards}, R.~T., \& {Manchester}, R.~N. 2006, \mnras, 369,
  655, \dodoi{10.1111/j.1365-2966.2006.10302.x}

\bibitem[{{Hotan} {et~al.}(2004{\natexlab{a}}){Hotan}, {Bailes}, \&
  {Ord}}]{2004MNRAS.355..941H}
{Hotan}, A.~W., {Bailes}, M., \& {Ord}, S.~M. 2004{\natexlab{a}}, \mnras, 355,
  941, \dodoi{10.1111/j.1365-2966.2004.08376.x}

\bibitem[{{Hotan} {et~al.}(2004{\natexlab{b}}){Hotan}, {van Straten}, \&
  {Manchester}}]{2004PASA...21..302H}
{Hotan}, A.~W., {van Straten}, W., \& {Manchester}, R.~N. 2004{\natexlab{b}},
  \pasa, 21, 302, \dodoi{10.1071/AS04022}

\bibitem[{{Jenet} {et~al.}(2001){Jenet}, {Anderson}, \&
  {Prince}}]{2001ApJ...546..394J}
{Jenet}, F.~A., {Anderson}, S.~B., \& {Prince}, T.~A. 2001, \apj, 546, 394,
  \dodoi{10.1086/318256}

\bibitem[{{Jiang} {et~al.}(2019){Jiang}, {Yue}, {Gan}, {Yao}, {Li}, {Pan},
  {Sun}, {Yu}, {Liu}, {Tang}, {Qian}, {Lu}, {Yan}, {Peng}, {Zhang}, {Wang},
  {Li}, \& {Li}}]{2019SCPMA..6259502J}
{Jiang}, P., {Yue}, Y., {Gan}, H., {et~al.} 2019, Science China Physics,
  Mechanics, and Astronomy, 62, 959502, \dodoi{10.1007/s11433-018-9376-1}

\bibitem[{{Johnston} \& {Romani}(2003)}]{2003ApJ...590L..95J}
{Johnston}, S., \& {Romani}, R.~W. 2003, \apjl, 590, L95,
  \dodoi{10.1086/376826}

\bibitem[{{Keith} {et~al.}(2013){Keith}, {Coles}, {Shannon}, {Hobbs},
  {Manchester}, {Bailes}, {Bhat}, {Burke-Spolaor}, {Champion}, {Chaudhary},
  {Hotan}, {Khoo}, {Kocz}, {Os{\l}owski}, {Ravi}, {Reynolds}, {Sarkissian},
  {van Straten}, \& {Yardley}}]{2013MNRAS.429.2161K}
{Keith}, M.~J., {Coles}, W., {Shannon}, R.~M., {et~al.} 2013, \mnras, 429,
  2161, \dodoi{10.1093/mnras/sts486}

\bibitem[{{Kerr} {et~al.}(2020){Kerr}, {Reardon}, {Hobbs}, {Shannon},
  {Manchester}, {Dai}, {Russell}, {Zhang}, {van Straten}, {Os{\l}owski},
  {Parthasarathy}, {Spiewak}, {Bailes}, {Bhat}, {Cameron}, {Coles}, {Dempsey},
  {Deng}, {Goncharov}, {Kaczmarek}, {Keith}, {Lasky}, {Lower}, {Preisig},
  {Sarkissian}, {Toomey}, {Wang}, {Wang}, {Zhang}, \&
  {Zhu}}]{2020PASA...37...20K}
{Kerr}, M., {Reardon}, D.~J., {Hobbs}, G., {et~al.} 2020, \pasa, 37, e020,
  \dodoi{10.1017/pasa.2020.11}

\bibitem[{{Kramer} \& {Champion}(2013)}]{2013CQGra..30v4009K}
{Kramer}, M., \& {Champion}, D.~J. 2013, Classical and Quantum Gravity, 30,
  224009, \dodoi{10.1088/0264-9381/30/22/224009}

\bibitem[{{Kramer} {et~al.}(1999){Kramer}, {Xilouris}, {Camilo}, {Nice},
  {Backer}, {Lange}, {Lorimer}, {Doroshenko}, \&
  {Sallmen}}]{1999ApJ...520..324K}
{Kramer}, M., {Xilouris}, K.~M., {Camilo}, F., {et~al.} 1999, \apj, 520, 324,
  \dodoi{10.1086/307449}

\bibitem[{{Lam} {et~al.}(2016){Lam}, {Cordes}, {Chatterjee}, {Arzoumanian},
  {Crowter}, {Demorest}, {Dolch}, {Ellis}, {Ferdman}, {Fonseca}, {Gonzalez},
  {Jones}, {Jones}, {Levin}, {Madison}, {McLaughlin}, {Nice}, {Pennucci},
  {Ransom}, {Siemens}, {Stairs}, {Stovall}, {Swiggum}, \&
  {Zhu}}]{2016ApJ...819..155L}
{Lam}, M.~T., {Cordes}, J.~M., {Chatterjee}, S., {et~al.} 2016, \apj, 819, 155,
  \dodoi{10.3847/0004-637X/819/2/155}

\bibitem[{{Li} \& {Pan}(2016)}]{li16}
{Li}, D., \& {Pan}, Z. 2016, Radio Science, 51, 1060,
  \dodoi{10.1002/2015RS005877}

\bibitem[{{Liu} {et~al.}(2012){Liu}, {Keane}, {Lee}, {Kramer}, {Cordes}, \&
  {Purver}}]{2012MNRAS.420..361L}
{Liu}, K., {Keane}, E.~F., {Lee}, K.~J., {et~al.} 2012, \mnras, 420, 361,
  \dodoi{10.1111/j.1365-2966.2011.20041.x}

\bibitem[{{Liu} {et~al.}(2015){Liu}, {Karuppusamy}, {Lee}, {Stappers},
  {Kramer}, {Smits}, {Purver}, {Janssen}, \& {Perrodin}}]{2015MNRAS.449.1158L}
{Liu}, K., {Karuppusamy}, R., {Lee}, K.~J., {et~al.} 2015, \mnras, 449, 1158,
  \dodoi{10.1093/mnras/stv397}

\bibitem[{{Liu} {et~al.}(2016){Liu}, {Bassa}, {Janssen}, {Karuppusamy},
  {McKee}, {Kramer}, {Lee}, {Perrodin}, {Purver}, {Sanidas}, {Smits},
  {Stappers}, {Weltevrede}, \& {Zhu}}]{2016MNRAS.463.3239L}
{Liu}, K., {Bassa}, C.~G., {Janssen}, G.~H., {et~al.} 2016, \mnras, 463, 3239,
  \dodoi{10.1093/mnras/stw2223}

\bibitem[{{Lyne} {et~al.}(2010){Lyne}, {Hobbs}, {Kramer}, {Stairs}, \&
  {Stappers}}]{2010Sci...329..408L}
{Lyne}, A., {Hobbs}, G., {Kramer}, M., {Stairs}, I., \& {Stappers}, B. 2010,
  Science, 329, 408, \dodoi{10.1126/science.1186683}

\bibitem[{{Mahajan} {et~al.}(2018){Mahajan}, {van Kerkwijk}, {Main}, \&
  {Pen}}]{2018ApJ...867L...2M}
{Mahajan}, N., {van Kerkwijk}, M.~H., {Main}, R., \& {Pen}, U.-L. 2018, \apjl,
  867, L2, \dodoi{10.3847/2041-8213/aae713}

\bibitem[{{Manchester} \& {IPTA}(2013)}]{2013CQGra..30v4010M}
{Manchester}, R.~N., \& {IPTA}. 2013, Classical and Quantum Gravity, 30,
  224010, \dodoi{10.1088/0264-9381/30/22/224010}

\bibitem[{{Manchester} {et~al.}(2013){Manchester}, {Hobbs}, {Bailes}, {Coles},
  {van Straten}, {Keith}, {Shannon}, {Bhat}, {Brown}, {Burke-Spolaor},
  {Champion}, {Chaudhary}, {Edwards}, {Hampson}, {Hotan}, {Jameson}, {Jenet},
  {Kesteven}, {Khoo}, {Kocz}, {Maciesiak}, {Oslowski}, {Ravi}, {Reynolds},
  {Sarkissian}, {Verbiest}, {Wen}, {Wilson}, {Yardley}, {Yan}, \&
  {You}}]{2013PASA...30...17M}
{Manchester}, R.~N., {Hobbs}, G., {Bailes}, M., {et~al.} 2013, \pasa, 30, e017,
  \dodoi{10.1017/pasa.2012.017}

\bibitem[{{Nan} {et~al.}(2011){Nan}, {Li}, {Jin}, {Wang}, {Zhu}, {Zhu},
  {Zhang}, {Yue}, \& {Qian}}]{2011IJMPD..20..989N}
{Nan}, R., {Li}, D., {Jin}, C., {et~al.} 2011, International Journal of Modern
  Physics D, 20, 989, \dodoi{10.1142/S0218271811019335}

\bibitem[{{Os{\l}owski} {et~al.}(2014){Os{\l}owski}, {van Straten}, {Bailes},
  {Jameson}, \& {Hobbs}}]{2014MNRAS.441.3148O}
{Os{\l}owski}, S., {van Straten}, W., {Bailes}, M., {Jameson}, A., \& {Hobbs},
  G. 2014, \mnras, 441, 3148, \dodoi{10.1093/mnras/stu804}

\bibitem[{{Os{\l}owski} {et~al.}(2011){Os{\l}owski}, {van Straten}, {Hobbs},
  {Bailes}, \& {Demorest}}]{2011MNRAS.418.1258O}
{Os{\l}owski}, S., {van Straten}, W., {Hobbs}, G.~B., {Bailes}, M., \&
  {Demorest}, P. 2011, \mnras, 418, 1258,
  \dodoi{10.1111/j.1365-2966.2011.19578.x}

\bibitem[{{Padmanabh} {et~al.}(2020){Padmanabh}, {Barr}, {Champion},
  {Karuppusamy}, {Kramer}, {Jessner}, \& {Lazarus}}]{2020MNRAS.tmp.2964P}
{Padmanabh}, P.~V., {Barr}, E.~D., {Champion}, D.~J., {et~al.} 2020, \mnras,
  \dodoi{10.1093/mnras/staa3174}

\bibitem[{{Shannon} \& {Cordes}(2012)}]{2012ApJ...761...64S}
{Shannon}, R.~M., \& {Cordes}, J.~M. 2012, \apj, 761, 64,
  \dodoi{10.1088/0004-637X/761/1/64}

\bibitem[{{Shannon} {et~al.}(2014){Shannon}, {Os{\l}owski}, {Dai}, {Bailes},
  {Hobbs}, {Manchester}, {van Straten}, {Raithel}, {Ravi}, {Toomey}, {Bhat},
  {Burke-Spolaor}, {Coles}, {Keith}, {Kerr}, {Levin}, {Sarkissian}, {Wang},
  {Wen}, \& {Zhu}}]{2014MNRAS.443.1463S}
{Shannon}, R.~M., {Os{\l}owski}, S., {Dai}, S., {et~al.} 2014, \mnras, 443,
  1463, \dodoi{10.1093/mnras/stu1213}

\bibitem[{{Shao} \& {You}(2017)}]{2017ChA&A..41..495S}
{Shao}, M., \& {You}, X.-p. 2017, \caa, 41, 495,
  \dodoi{10.1016/j.chinastron.2017.11.002}

\bibitem[{{Staelin} \& {Reifenstein}(1968)}]{1968Sci...162.1481S}
{Staelin}, D.~H., \& {Reifenstein}, Edward~C., I. 1968, Science, 162, 1481,
  \dodoi{10.1126/science.162.3861.1481}

\bibitem[{{Stairs} {et~al.}(2019){Stairs}, {Lyne}, {Kramer}, {Stappers}, {van
  Leeuwen}, {Tung}, {Manchester}, {Hobbs}, {Lorimer}, \&
  {Melatos}}]{2019MNRAS.485.3230S}
{Stairs}, I.~H., {Lyne}, A.~G., {Kramer}, M., {et~al.} 2019, \mnras, 485, 3230,
  \dodoi{10.1093/mnras/stz647}

\bibitem[{{van Straten} \& {Bailes}(2011)}]{2011PASA...28....1V}
{van Straten}, W., \& {Bailes}, M. 2011, \pasa, 28, 1, \dodoi{10.1071/AS10021}

\bibitem[{{Wang} {et~al.}(2007){Wang}, {Manchester}, \&
  {Johnston}}]{2007MNRAS.377.1383W}
{Wang}, N., {Manchester}, R.~N., \& {Johnston}, S. 2007, \mnras, 377, 1383,
  \dodoi{10.1111/j.1365-2966.2007.11703.x}

\bibitem[{{Wang} {et~al.}(2020){Wang}, {Wang}, {Hobbs}, {Zhang}, {Shannon},
  {Dai}, {Hollow}, {Kerr}, {Ravi}, {Wang}, \& {Zhang}}]{2020ApJ...897....8W}
{Wang}, S.~Q., {Wang}, J.~B., {Hobbs}, G., {et~al.} 2020, \apj, 897, 8,
  \dodoi{10.3847/1538-4357/ab9302}

\bibitem[{{Weltevrede} {et~al.}(2006){Weltevrede}, {Edwards}, \&
  {Stappers}}]{2006A&A...445..243W}
{Weltevrede}, P., {Edwards}, R.~T., \& {Stappers}, B.~W. 2006, \aap, 445, 243,
  \dodoi{10.1051/0004-6361:20053088}

\end{thebibliography}
\bibliographystyle{aasjournal}



\end{document}